\documentclass[]{article}
\usepackage{amsmath,amssymb}
\usepackage{hyperref}
\hypersetup{colorlinks,
            linkcolor=[HTML]{009DE0},
            anchorcolor=[HTML]{009DE0},
            citecolor=[HTML]{009DE0},
            urlcolor=[HTML]{009DE0}}
\usepackage[T1]{fontenc}
\usepackage[utf8]{inputenc}
\usepackage{xcolor}
\usepackage[margin=1in]{geometry}
\usepackage{fancyvrb}

\DefineVerbatimEnvironment{Highlighting}{Verbatim}{commandchars=\\\{\}}
\usepackage{framed}
\definecolor{shadecolor}{RGB}{248,248,248}
\newenvironment{Shaded}{\begin{snugshade}}{\end{snugshade}}

\newcommand{\AttributeTok}[1]{\textcolor[rgb]{0.13,0.29,0.53}{#1}}

\newcommand{\DecValTok}[1]{\textcolor[rgb]{0.00,0.00,0.81}{#1}}

\newcommand{\FloatTok}[1]{\textcolor[rgb]{0.00,0.00,0.81}{#1}}
\newcommand{\FunctionTok}[1]{\textcolor[rgb]{0.13,0.29,0.53}{\texttt{#1}}}

\newcommand{\NormalTok}[1]{#1}

\newcommand{\OtherTok}[1]{\textcolor[rgb]{0.56,0.35,0.01}{#1}}

\newcommand{\SpecialCharTok}[1]{\textcolor[rgb]{0.81,0.36,0.00}{\texttt{#1}}}

\newcommand{\StringTok}[1]{\textcolor[rgb]{0.31,0.60,0.02}{#1}}

\usepackage{graphicx}
\usepackage{subcaption}

\usepackage[round]{natbib}
\title{AIDA: Analysis of Interval Data in R}
\author{
  Catarina P. Loureiro$^{*}$ \\
  \small CEMAT \& Department of Mathematics, Instituto Superior Técnico, Universidade de Lisboa \\
  \small $^{*}$Corresponding author: \href{mailto:catarinapadrela@tecnico.ulisboa.pt}{\texttt{catarinapadrela@tecnico.ulisboa.pt}}
  \and
  M. Rosário Oliveira \\
  \small CEMAT \& Department of Mathematics, Instituto Superior Técnico, Universidade de Lisboa \\
  \small \href{mailto:rosario.oliveira@tecnico.ulisboa.pt}{\texttt{rosario.oliveira@tecnico.ulisboa.pt}}
  \and
  Paula Brito \\
  \small Faculdade de Economia e Gestão, Universidade do Porto \& LIAAD-INESC TEC \\
  \small \href{mailto:mpbrito@fep.up.pt}{\texttt{mpbrito@fep.up.pt}}
}
\date{}

\begin{document}
\maketitle
\begin{abstract}
\texttt{AIDA} is an R package that provides tools for the analysis of interval-valued data, including construction, visualization, robust estimation, and explainable outlier detection. The package is built around the \texttt{intData} class and is designed to support methodological research and applied workflows involving interval-valued data. \texttt{AIDA} implements methods for symbolic covariance estimation based on the Mallows distance, robust estimation via the Interval Minimum Covariance Determinant estimator, and explainable outlier detection using Shapley values. The package also includes visualization tools to support interpretation of the results and several interval-valued datasets.
\end{abstract}

\section{Introduction}\label{sec:introduction}

Symbolic Data Analysis (SDA) \citep{Diday} is a growing field of statistics dedicated to the analysis of complex data, such as lists, intervals or distributions, leveraging structured or aggregated data representations rather than conventional single-valued observations \citep{SDAoverview}. This allows the inherent variability of the data to be preserved, which would otherwise be lost. Symbolic data can appear naturally when collections of individual observations (microdata) are aggregated into macrodata, or when only summary information is available. Interval-valued data, in particular, constitute a common and useful type of symbolic data, providing a structured representation of an underlying variability pattern. This ontic perspective \citep{COUSO2014} is central to the analysis of interval data and motivates the development of methods that account for both the location and the internal dispersion of each observation. Interval-valued data arise in a wide range of applications, such as climate records, economic indicators, environmental monitoring, and aggregated consumer behavior, among others. For example, a temperature may be recorded as a daily range rather than a single value, or a set of measurements may be represented by an interval to preserve privacy or reduce dimensionality. In such settings, the analyst works with a data object that contains more information than a point, but less than a full microdata sample.

Several R packages provide tools for symbolic data analysis, including \texttt{MAINT.Data} \citep{MAINT.Data.article} (modeling multivariate interval data by Normal and Skew-Normal distributions, including maximum likelihood estimation, (M)ANOVA, Robust Discriminant Analysis, Model-based Clustering, and outlier detection); \texttt{RSDA} \citep{RSDA} (automatic classification and linear models); \texttt{dataSDA} \citep{dataSDA} (benchmark datasets and foundational tools); \texttt{ggInterval} \citep{ggInterval} (extends \texttt{ggplot2} for visualizing interval-valued data); \texttt{HistDAWass} \citep{HistDAWass} (unsupervised classification, least squares regression, and tools for histogram-valued data and for histogram time series); \texttt{mdsOpt} \citep{mdsOpt} (selection of the optimal multidimensional scaling (MDS) procedure for interval-valued data); \texttt{iRegression} \citep{iRegression} (regression methods for interval-valued variables); and \texttt{symbolicDA} \citep{symbolicDA} (general multivariate data analysis, including estimation and visualization methods, MDS, clustering, decision trees, kernel discriminant analysis, among others). However, these packages are often specialized for specific tasks or data representations, and they do not provide a comprehensive foundation for the analysis of interval-valued data that takes into account information from both microdata and macrodata. \texttt{AIDA} \citep{AIDA} was developed to provide an integrated framework for such a workflow, including data construction, robust estimation, explainable outlier detection, and graphical exploration.

\texttt{AIDA} is built around the \texttt{intData} S4 class, which represents interval-valued observations while retaining the information required to model their internal variability. The package supports the construction of interval-valued datasets from raw microdata or aggregated summaries, descriptive statistics and symbolic covariance matrices, robust estimation of location and scatter, outlier detection using robust Interval-Mahalanobis distances, and explainability through Shapley values and interaction indices. Its methodological foundation is the model linking microdata and macrodata \citep{Oliveira2022}, which provides a latent representation of interval data and motivates definitions of barycenter and covariance based on the Mallows distance \citep{Mallows1972, Irpino2015, oliveira2025}. For robust analysis, \texttt{AIDA} implements the Interval Minimum Covariance Determinant (IMCD) estimator \citep{loureiro2026}, together with robust Interval-Mahalanobis distances for outlier detection. The package further provides Shapley value-based methods for decomposing outlyingness into variable-specific contributions \citep{loureiro2026b}, allowing users to investigate each variable's importance to the anomalous behavior. The package also includes visualization methods and datasets to aid exploratory analysis and interpretation.

In the sections that follow, we introduce the interval-valued data framework and relevant theoretical background (\hyperref[sec:interval-valued-data]{Section 2}), describe the main statistical tools implemented in \texttt{AIDA} (\hyperref[sec:package-structure]{Section 3}), and illustrate their use on real interval-valued datasets (\hyperref[sec:examples]{Section 4}). Finally, \hyperref[sec:summary]{Section 5} summarizes the package's contributions.

\section{Interval-valued data}\label{sec:interval-valued-data}

Interval-valued data represent each observation as a \(p\)-dimensional hyperrectangle, with each variable described by an interval rather than a single value. The set of \(p\)-dimensional hyperrectangles is denoted by \(\mathbb{IR}^p = \{([a_1,b_1],\dots,[a_p,b_p])^\top:\; a_j,b_j \in \mathbb{R},\; a_j \leq b_j, j=1,\dots,p \}\), with \(p\) \(\in \mathbb{N}\). Each real-valued interval \([a,b]\in\mathbb{IR}\) can be equivalently represented by its center, \(c=(a+b)/2\in\mathbb{R}\), and its range, \(r=b-a\in\mathbb{R}_0^+\).

In SDA, a symbolic interval is often interpreted as combining two types of information: \emph{macrodata}, the real-valued interval itself (through its center and range); \emph{microdata}, the observed values that lie within the real-valued interval. This symbolic interval can be denoted by \(x=(c,r,F_U)\) \citep{oliveira2025}, where \(F_U\) is the distribution function of a real-valued latent random variable, \(U\), with support \([-1,1]\) (which corresponds to the microdata normalized to \([-1,1]\)). When the microdata are not available, it is common to assume a continuous uniform distribution within the interval, which is the default behavior in \texttt{AIDA}. When the microdata are available, the \texttt{AIDA} package can be used to infer a richer latent distribution. This motivates the \texttt{intData} class, which stores, among other information, the intervals' centers and ranges, and the parameters of the latent variables' distribution.

In this framework, the real-valued random vectors of centers and ranges can be denoted by \(\boldsymbol{\mathcal{C}} = (C_1,\ldots,C_p)^\top\) and \(\boldsymbol{\mathcal{R}} = (R_1,\ldots,R_p)^\top\), respectively, such that \(\mathbb{P}(\boldsymbol{\mathcal{R}} \geq \boldsymbol{0}) = 1\). In addition, let \(\boldsymbol{U} = (U_1,\ldots,U_p)^\top\) be the real-valued latent random vector of independent and absolutely continuous random variables with support \(\left[-1,1\right]^p\) and joint distribution function \(F_{\boldsymbol{U}}\). Then, an interval-valued random vector is denoted by \(\boldsymbol{\mathcal{X}} = (X_1,\ldots,X_p)^\top\), and can be characterized as \(\boldsymbol{\mathcal{X}}=(\boldsymbol{\mathcal{C}},\boldsymbol{\mathcal{R}},F_{\boldsymbol{U}})\). Assuming that \(\boldsymbol{V} = (V_1,\ldots,V_p)^\top\in\mathbb{R}^p\) represents the microdata, \citet{Oliveira2022} defines the model linking the microdata and macrodata as \(V_j=C_j+U_jR_j/2\), if \(\mathbb{P}(R_j=0)=0\), and \(V_j=C_j\), if \(\mathbb{P}(R_j=0)=1\) and \(\mathbb{P}(U_j=0)=1\).

\subsubsection*{Mallows distance}\label{sec:mallows}

A key ingredient in the analysis of interval-valued data is a notion of distance. A widely used choice is the Mallows distance \citep{Mallows1972, Irpino2015, oliveira2025}, which compares intervals by taking into account both their location and their internal variability. For two intervals, \(x_1=((c_1,r_1)^\top,F_{1})\) and \(x_2=((c_2,r_2)^\top,F_{2})\), with \(F_j\) and \(F_{j}^{-1}\) the distribution function (with finite second moment) and quantile function of the microdata, respectively, for \(j=1,2\), the Mallows distance between them is defined as \(d_\mathrm{M}^2(x_1, x_2) = \int_0^1 (F_{1}^{-1}(t) - F_{2}^{-1}(t))^2 \, dt\). Thus, this distance can be interpreted as the integral of the \(L_2\) distance between the quantile functions of the microdata distributions. In practice, it can be expressed in terms of centers and ranges, which makes it easy to compute and interpret. Let \(\boldsymbol{x}_i=(\boldsymbol{c}_i,\boldsymbol{r}_i,F_{\boldsymbol{U}})\) be the \(i\)-th realization of an interval-valued random vector \(\boldsymbol{\mathcal{X}}\), with centers \(\boldsymbol{c}_i=(c_{i1},\dots,c_{ip})^\top\in\mathbb{R}^p\), and ranges \(\boldsymbol{r}_i=(r_{i1},\dots,r_{ip})^\top\in(\mathbb{R}^{+}_0)^p\), for \(i=1,2\). Assuming \(\boldsymbol{U}\) has finite second moments and \(\mathrm{diag}(\boldsymbol{v})\) denotes the diagonal matrix whose main diagonal is the vector \(\boldsymbol{v}\), the squared Mallows distance between these intervals \citep{oliveira2025} can be computed by
\begin{equation}
    d_\mathrm{M}^2(\boldsymbol{x}_1, \boldsymbol{x}_2) = (\boldsymbol{c}_1-\boldsymbol{c}_2)^\top(\boldsymbol{c}_1-\boldsymbol{c}_2) + (\boldsymbol{r}_1-\boldsymbol{r}_2)^\top\boldsymbol{\Delta}(\boldsymbol{r}_1-\boldsymbol{r}_2) + (\boldsymbol{c}_1-\boldsymbol{c}_2)^\top\boldsymbol{\Psi}(\boldsymbol{r}_1-\boldsymbol{r}_2),
     \label{eq:mallows}
\end{equation}
where \(\boldsymbol{\Delta}=\mathrm{diag}(\delta_1,\dots,\delta_p)\), with \(\delta_j=\mathbb{E}(U_j^2)/4\), \(\boldsymbol{\Psi}=\mathrm{diag}(\mathbb{E}(U_1),\dots,\mathbb{E}(U_p))\). This is the expression used to implement the Mallows distance in the \texttt{AIDA} package.

\textbf{Example 1.} \emph{For the case when the latent variables are uniformly distributed, \(U_j\sim\mathrm{Unif}(-1,1)\), we have \(\delta_j=1/12\) and \(\mathbb{E}(U_j)=0\), so the squared Mallows distance reduces to \(d_\mathrm{M}^2(\boldsymbol{x}_1, \boldsymbol{x}_2) = (\boldsymbol{c}_1-\boldsymbol{c}_2)^\top(\boldsymbol{c}_1-\boldsymbol{c}_2) + \frac{1}{12}(\boldsymbol{r}_1-\boldsymbol{r}_2)^\top(\boldsymbol{r}_1-\boldsymbol{r}_2)\).}

\subsection{Robust estimation of the symbolic covariance matrix}\label{sec:robust-cov-matrix}

Using the Mallows distance, it is possible to define summary statistics for interval-valued data. A natural notion of mean is given by the barycenter, which is the interval that minimizes the average squared Mallows distance to the data \citep{Irpino2015, Irpino2006}. Based on the model linking microdata and macrodata and assuming that the centers and ranges have finite expected values, \(\boldsymbol{\mu}_C\) and \(\boldsymbol{\mu}_R\), respectively, and that the latent random vector \(\boldsymbol{U}\) is independent of \(\boldsymbol{\mathcal{X}}\), the population barycenter has a simple form. In fact, it is obtained by averaging the centers and the ranges separately resulting in the interval \(\boldsymbol{\mu}_B=(\boldsymbol{\mu}_C, \boldsymbol{\mu}_R, F_{\boldsymbol{U}})\) \citep{oliveira2025}.

A symbolic covariance matrix can then be derived from the same ingredients, extending the classical covariance to interval-valued data. Let \(\boldsymbol{\Sigma}_{CC}\) and \(\boldsymbol{\Sigma}_{RR}\) be the covariance matrices of the centers and ranges, respectively, and \(\boldsymbol{\Sigma}_{CR}\) the covariance matrix between centers and ranges. The symbolic covariance matrix can be written as \citep{oliveira2025}
\begin{equation}
    \boldsymbol{\Sigma}_B=\boldsymbol{\Sigma}_{CC}+\dfrac{1}{4}\boldsymbol{\mathfrak{E}}_{UU}\bullet\boldsymbol{\Sigma}_{RR}+\dfrac{1}{2}\boldsymbol{\Sigma}_{CR}\boldsymbol{\Psi}+\dfrac{1}{2}\boldsymbol{\Psi}\boldsymbol{\Sigma}_{RC},
    \label{eq:covmatrix}
\end{equation}
where \([\boldsymbol{\mathfrak{E}}_{UU}]_{jj}=\mathbb{E}(U_j^2)\), \([\boldsymbol{\mathfrak{E}}_{UU}]_{j\ell}=\mathcal{E}(U_j,U_\ell)\), \(j\neq \ell\), with \(\mathcal{E}(U_j,U_\ell)=\mathbb{E}(F_{U_j}^{-1}(T) F_{U_\ell}^{-1}(T))=\int_0^1 F_{U_j}^{-1}(t) F_{U_\ell}^{-1}(t) \, dt\), \(T\sim\mathrm{Unif}(0,1)\), \(\boldsymbol{\Psi}=\mathrm{diag}(\mathbb{E}(U_1),\dots,\mathbb{E}(U_p))\), \(\bullet\) denotes the Schur or entrywise product of matrices, and \([\boldsymbol{A}]_{j\ell}\) the entry \((j,\ell)\) of matrix \(\boldsymbol{A}\). This covariance matrix provides a measure of dispersion and association adapted to interval-valued data.

\textbf{Example 2.} \emph{When the latent variables are uniformly distributed, \(U_j\sim\mathrm{Unif}(-1,1)\), we have \([\boldsymbol{\mathfrak{E}}_{UU}]_{jj}=\mathbb{E}(U_j^2)=4 \delta_j\), \([\boldsymbol{\mathfrak{E}}_{UU}]_{j\ell}=0\), and \(\boldsymbol{\Psi}=\boldsymbol{0}\), and the symbolic covariance matrix reduces to \(\boldsymbol{\Sigma}_B=\boldsymbol{\Sigma}_{CC}+\frac{1}{12}\boldsymbol{\Sigma}_{RR}\).}

As is well known, sample estimators of the barycenter and the covariance matrix are sensitive to the presence of atypical observations (outliers), whose presence is very common in real-world datasets. This is an issue that extends to the context of interval-valued data, potentially leading to biased estimates. To address this issue, the IMCD algorithm \citep{loureiro2026} extends the FastMCD algorithm \citep{fastMCD} to interval-valued data, providing robust estimates of both the barycenter and the symbolic covariance matrix. The method identifies a subset of observations that minimizes the determinant of the covariance matrix, thereby limiting the influence of outliers. A reweighting step based on robust distances and a cutoff value is then applied to improve efficiency. These procedures are implemented in the \texttt{AIDA} package, notably through functions for computing the symbolic covariance matrix and implementing the IMCD algorithm.

\subsection{Outlier detection based on robust Interval-Mahalanobis distance}\label{sec:outlier-detection}

A central tool in Multivariate Analysis is the Mahalanobis distance. For interval-valued data, this concept can be extended to the Interval-Mahalanobis distance \citep{loureiro2026}, which accounts for the specific structure of intervals. Let \(\boldsymbol{x}=(\boldsymbol{c},\boldsymbol{r},F_{\boldsymbol{U}})\), with \(\boldsymbol{c}=(c_1,\dots,c_p)^\top\in\mathbb{R}^p\), \(\boldsymbol{r}=(r_1,\dots,r_p)^\top\in(\mathbb{R}^{+}_0)^p\), and \(\boldsymbol{U}=(U_1,\dots,U_p)^\top\) a latent random vector of independent random variables supported on \(\left[-1,1\right]^p\) with joint distribution function \(F_{\boldsymbol{U}}\). Then, the squared Interval-Mahalanobis distance can be written as
\begin{equation}
  \begin{aligned}
      d_\mathrm{IMah}^2(\boldsymbol{x})&=(\boldsymbol{c}-\boldsymbol{\mu}_C)^\top\boldsymbol{\Sigma}_B^{-1}(\boldsymbol{c}-\boldsymbol{\mu}_C)+\frac{1}{4}(\boldsymbol{r}-\boldsymbol{\mu}_R)^\top\left(\boldsymbol{\mathfrak{E}}_{UU}\bullet\boldsymbol{\Sigma}_B^{-1}\right)(\boldsymbol{r}-\boldsymbol{\mu}_R)\\
      &\quad +(\boldsymbol{c}-\boldsymbol{\mu}_C)^\top\boldsymbol{\Sigma}_B^{-1}\boldsymbol{\Psi}(\boldsymbol{r}-\boldsymbol{\mu}_R),
  \end{aligned}
  \label{eq:d2mah}
\end{equation}
where \(\boldsymbol{\mu}_C\) and \(\boldsymbol{\mu}_R\) denote the mean vectors of the centers and ranges, respectively, and \(\boldsymbol{\Sigma}_B\) the symbolic covariance matrix. The Mallows distance \eqref{eq:mallows} is a special case of the Interval-Mahalanobis distance with the covariance matrix replaced by the identity matrix.

In practice, this distance measures how far an observation is from the barycenter of the data, taking into account the variability described by the symbolic covariance matrix. A robust version of this distance can be obtained by replacing the classical estimates of location and covariance with the robust estimates provided by the IMCD algorithm. The resulting robust Interval-Mahalanobis distance provides a measure of outlyingness: observations that are far from the robust center of the data are more likely to be outliers. In this framework, an observation is flagged as an outlier if its distance exceeds a given cutoff value.

Since the distribution of this distance is not known, cutoff values can be obtained using a nonparametric approach \citep{loureiro2026} based on the concept of farness \citep{classmap}. This approach involves transforming the squared distances to obtain approximately normally distributed scores. The transformation includes: a robust standardization using the median and median absolute deviation (MAD), a robust Yeo-Johnson transformation \citep{yeojohnson2000, Raymaekers2024} to reduce skewness, and a second standardization step. The transformed values are then converted into outlier scores using the cumulative distribution function of the standard normal distribution. A common choice is to flag observations with scores above 0.95 as outliers, corresponding to an approximate \(5\%\) false positive rate. Package \texttt{AIDA} implements the computation of the Interval-Mahalanobis distance and the outlier detection rule (with different choices available for estimating the cutoff value).

\subsection{Explainability with Shapley values}\label{sec:explainability}

As machine learning models become increasingly complex, there is a growing need for methods that help interpret their predictions. This has motivated the development of explainable artificial intelligence (XAI), which aims to make model outputs more transparent. Popular approaches include LIME and Shapley values (see, e.g., \citet{XAI2020}).

Shapley values, originally introduced by \citep{Shapley1953} in cooperative game theory, provide a principled way to attribute a total quantity to individual players. Let a set of players be \(P=\{1,\dots,p\}\) and a value function \(v\) assign a worth \(v(S)\) to each subset \(S\subseteq P\). Let \(|\cdot|\) denote set cardinality, then the Shapley value of player \(j\) is defined as
\begin{equation}
    \phi_j(v)=\sum_{S\subseteq P\setminus\{j\}}\frac{|S|!(p-|S|-1)!}{p!}\left(v(S\cup\{j\})-v(S)\right),
    \label{eq:shapleygame}
\end{equation}
i.e., the average marginal contribution of player \(j\) across all subsets that do not include \(j\). Shapley values are uniquely characterized by a set of desirable properties, including efficiency (the contributions sum to the total value), among others \citep{Young1985}.

In machine learning, variables play the role of players and Shapley values quantify their contributions to a model output \citep{SHAP2017}. In outlier detection, they can be used to decompose an outlyingness measure into variable-wise contributions \citep{shapley}. Using the squared Interval-Mahalanobis distance as the value function, the Shapley value of variable \(j\) then measures its average contribution to the total distance. A closed-form expression exists and can be written componentwise as a sum of center, range, and interaction terms \citep{loureiro2026b}. Here, we rely on the vector form for the Shapley values of an interval-valued observation \(\boldsymbol{x}\), \(\boldsymbol{\phi}(\boldsymbol{x})=(\phi_1(\boldsymbol{x}),\dots,\phi_p(\boldsymbol{x}))^\top\), given by
\begin{equation}
    \begin{aligned}
        \boldsymbol{\phi}(\boldsymbol{x})
        &=(\boldsymbol{c}-\boldsymbol{\mu}_C)\bullet\left[\boldsymbol{\Sigma}_B^{-1}(\boldsymbol{c}-\boldsymbol{\mu}_C)\right]+\frac{1}{4}(\boldsymbol{r}-\boldsymbol{\mu}_R)\bullet\left[\left(\boldsymbol{\mathfrak{E}}_{UU}\bullet\boldsymbol{\Sigma}_B^{-1}\right)(\boldsymbol{r}-\boldsymbol{\mu}_R)\right]\\
        &\quad+\frac{1}{2}(\boldsymbol{c}-\boldsymbol{\mu}_C)\bullet\left[\boldsymbol{\Sigma}_B^{-1}\boldsymbol{\Psi}(\boldsymbol{r}-\boldsymbol{\mu}_R)\right]+\frac{1}{2}(\boldsymbol{r}-\boldsymbol{\mu}_R)\bullet\left[\boldsymbol{\Psi}\boldsymbol{\Sigma}_B^{-1}(\boldsymbol{c}-\boldsymbol{\mu}_C)\right].
    \end{aligned}
    \label{eq:shapleyvector}
  \end{equation}

This expression is of linear computational cost, avoiding the usual combinatorial cost. In addition, by the efficiency property, we have \(\sum_{j=1}^p\phi_j(\boldsymbol{x})=d_\mathrm{IMah}^2(\boldsymbol{x})\), so the Shapley values provide an additive decomposition of the total outlyingness measure. Large positive values indicate variables driving the outlyingness, while negative values indicate a stabilizing effect. An additional level of decomposition can naturally be considered, separating the Shapley value expression into contributions from centers, ranges, and interaction between them. Moreover, Shapley values are closely related to the notion of cellwise outliers \citep{Raymaekers_Rousseeuw_2021}, as they allow the detection of observations that behave anomalously in certain variables even when the overall observation is not flagged as an outlier.

\subsubsection*{Shapley interaction index}\label{sec:shapley-inter-index}

The Shapley interaction index \citep{Grabisch1999, Fujimoto2006} further extends Shapley values to quantify joint effects between variables. For a subset \(Q\subseteq P\), it is defined as
\begin{equation}
    \Phi_Q(v)=\sum_{T\subseteq P\setminus Q}\frac{|T|!(p-|T|-|Q|)!}{(p-|Q|+1)!}\Delta_Q\, v(T),
    \label{eq:shapleyintergame}
\end{equation}
where \(\Delta_Q v(T)=\sum_{S\subseteq Q}(-1)^{|Q|-|S|}v(T\cup S)\). In our setting, we consider pairwise interactions (\(Q=\{j,\ell\}\)) for the squared Interval-Mahalanobis distance \(d_{\mathrm{IMah}}^2(\boldsymbol{x})\). The quantity \(\Phi_{j\ell}(\boldsymbol{x})\) measures the joint contribution of variables \(j\) and \(\ell\) to the total outlyingness.

The Shapley interaction indexes of an interval-valued observation \(\boldsymbol{x}\) for the pairs \((j,\ell)\), \(j,\ell=1,\dots,p\), can be collected in the \(p\times p\) matrix \(\boldsymbol{\Phi}(\boldsymbol{x})\), where \([\boldsymbol{\Phi}]_{j\ell}=\Phi_{j\ell}\), and given by \citep{loureiro2026b}
\begin{equation}
      \begin{aligned}
          \boldsymbol{\Phi}(\boldsymbol{x})&=2(\boldsymbol{c}-\boldsymbol{\mu}_C)(\boldsymbol{c}-\boldsymbol{\mu}_C)^\top\bullet\boldsymbol{\Sigma}_B^{-1}+\frac{1}{2}(\boldsymbol{r}-\boldsymbol{\mu}_R)(\boldsymbol{r}-\boldsymbol{\mu}_R)^\top\bullet\boldsymbol{\mathfrak{E}}_{UU}\bullet\boldsymbol{\Sigma}_B^{-1}+\\
          &\quad+(\boldsymbol{c}-\boldsymbol{\mu}_C)(\boldsymbol{r}-\boldsymbol{\mu}_R)^\top\bullet\boldsymbol{\Sigma}_B^{-1}\boldsymbol{\Psi}+(\boldsymbol{r}-\boldsymbol{\mu}_R)(\boldsymbol{c}-\boldsymbol{\mu}_C)^\top\bullet\boldsymbol{\Psi}\boldsymbol{\Sigma}_B^{-1}-\mathrm{diag}(\boldsymbol{\phi}(\boldsymbol{x})).
      \end{aligned}
    \label{eq:shapleyinter}
  \end{equation}
By construction, we have \(d_\mathrm{IMah}^2(\boldsymbol{x})=\sum_{j=1}^p\sum_{\ell=1}^p \Phi_{j\ell}(\boldsymbol{x})\), so the interaction index provides a full decomposition of the total outlyingness. Off-diagonal elements capture pairwise interactions, while diagonal elements represent individual effects after accounting for interactions. Thus, this representation reveals variable interactions in the outlyingness structure.

The three levels of explainability are implemented in \texttt{AIDA}. In addition, the package provides visualization tools to support the interpretation of the results, including bar plots for Shapley values and heatmaps for Shapley interaction indexes, among others. These tools allow users to explore the contributions of individual variables and their interactions to the overall outlyingness of interval-valued observations.

\section{Package structure and implementation}\label{sec:package-structure}

The package is organized around the central data structure for interval-valued data complemented by a set of analytical methods and visualization tools designed to facilitate exploratory analysis and, in particular, outlier detection.

\subsection{\texorpdfstring{The \texttt{intData} class}{The intData class}}\label{sec:intData}

The core data structure of the package is the \texttt{intData} class, an S4 class specifically designed to represent interval-valued data in a coherent and computationally efficient manner. The design of the \texttt{intData} class is partially inspired by the \texttt{IData} class from the \texttt{MAINT.Data} package, extending it with additional flexibility in data input formats and an explicit latent variable framework tailored to the methodological developments implemented in this package. Standard data structures in R, such as data frames or matrices, are not designed to explicitly encode the layered nature of interval-valued data, motivating the need for a dedicated class. The \texttt{intData} class adopts a center-range parameterization, storing interval-valued observations through two complementary components: a data frame of centers and a data frame of ranges (slots \texttt{Centers} and \texttt{Ranges}). This representation is particularly convenient for statistical analysis, as many operations can be naturally expressed in terms of these two quantities while maintaining a clear interpretation.

Beyond the basic interval structure, the class incorporates information about the underlying microdata through a latent variable framework. Specifically, each interval is assumed to arise from an aggregation of latent microdata, whose distributional properties are encoded via three slots: \texttt{LatentCase}, \texttt{LatentDist}, and \texttt{LatentParam}. The \texttt{LatentCase} slot distinguishes between three scenarios: identically distributed and symmetric variables (\texttt{"U\_id\_symmetric"}), identically distributed and asymmetric variables (\texttt{"U\_id"}), and a fully general case (\texttt{"General"}). The \texttt{LatentDist} specifies the assumed distribution(s) (uniform, triangular, truncated normal, inverse triangular, beta, kernel-based, or degenerated). The corresponding parameters are stored in \texttt{LatentParam}, allowing both predefined and data-driven specifications. The \texttt{LatentParam} is defined according to the \texttt{LatentCase}: for \texttt{"U\_id\_symmetric"}, we have \(\delta=\mathbb{E}(U^2)/4\); for \texttt{"U\_id"}, we have \(\delta=\mathbb{E}(U^2)/4\) and \(\mathbb{E}(U)\); and for \texttt{"General"}, we have \(\boldsymbol{\mathfrak{E}}_{UU}\) and \(\boldsymbol{\Psi}\). This design enables the integration of distributional assumptions directly into the data object, facilitating downstream methods that rely on these properties. Additional slots store metadata and structural information, including observation and variable names (\texttt{ObsNames} and \texttt{VarNames}), the number of observations and interval variables (\texttt{NObs} and \texttt{NIVar}), and the number of underlying micro-units aggregated into each interval (\texttt{NbMicroUnits}). If this information is unavailable, \texttt{NbMicroUnits} is set to an empty integer vector. These slots ensure that the object remains self-contained and suitable for both analysis and interpretation.

Objects of class \texttt{intData} are created using a dedicated constructor function (\texttt{intData()}) that accepts interval data in several common formats. In particular, the input can be provided as matrices or data frames with either lower and upper bounds or centers and ranges, with flexible ordering conventions (variable-by-variable or grouped formats). This is translated by the \texttt{Seq} argument with options: \texttt{"AllLb\_AllUb"}/\texttt{"AllCen\_AllRng"}, all lower bounds/centers followed by all upper bounds/ranges, in the same variable order; \texttt{"LbUb\_VarbyVar"}/\texttt{"CenRng\_VarbyVar"}, lower bounds/centers followed by upper bounds/ranges, variable by variable. The constructor automatically converts the input into the internal center-range representation and performs validation checks to ensure consistency, such as verifying that the number of columns is even and that interval widths are well-defined. If the parameters of the latent distributions are not explicitly provided, function \texttt{get\_latent\_param()} is called to estimate them from the data or preset assumptions, which is further discussed below. This allows users to seamlessly transition from raw aggregated data to a fully specified interval-valued object without requiring manual parameter specification.

In addition to data storage, the \texttt{intData} class provides basic methods for interaction and inspection. Standard methods for subsetting, printing, and summarizing are implemented, ensuring an intuitive and user-friendly interface within the R environment. In particular, a dedicated \texttt{summary} method is provided for objects of class \texttt{intData}, returning an object of class \texttt{summaryintData} that separately summarizes interval centers and ranges. This allows users to quickly inspect the main distributional characteristics of the data in a structured and interpretable way. Overall, the \texttt{intData} class serves as the foundation of the package, enabling a unified workflow in which interval-valued data can be created, manipulated, and analyzed using a consistent interface.

\subsubsection*{Aggregation of microdata into interval data}\label{sec:aggregate-microdata}

In many practical applications, interval-valued data do not arise directly but are obtained through the aggregation of underlying microdata. To support this workflow, the package provides the function \texttt{micro2intData()}, which transforms classical datasets into objects of class \texttt{intData} by grouping individual observations. The function takes as input a data frame of numeric microdata and a grouping factor defining how observations should be aggregated. By default, the intervals are defined using the minimum and maximum values within each group. Alternatively, more robust intervals can be obtained by specifying lower and upper quantiles, allowing the user to mitigate the influence of extreme observations. The implementation of this function is adapted from aggregation procedures available in the \texttt{MAINT.Data} package, extending them with additional flexibility in aggregation criteria and tighter integration with the latent variable modeling framework.

The aggregation procedure includes several validation and preprocessing steps to ensure the consistency of the resulting interval data. Observations with non-finite values are removed prior to aggregation, and groups containing insufficient information (e.g., variables with only missing values) are automatically discarded. The function also tracks the number of micro-observations contributing to each interval, storing this information within the resulting \texttt{intData} object. In addition to constructing interval bounds, the function integrates seamlessly with the latent variable framework underlying the \texttt{intData} class. When required, latent microdata representations are derived from the aggregated data and used to estimate distributional parameters under the specified assumptions (through function \texttt{get\_latent\_param()}). In particular, if the \texttt{LatentCase} is set to \texttt{"General"} and nothing is specified about the \texttt{LatentDist} (or it is set to \texttt{"KDE"}), Kernel Density Estimation (KDE) is used through package \texttt{kde1d} \citep{kde1d} to estimate the parameters. This allows the user to move directly from raw microdata to a fully specified interval-valued representation, including both descriptive bounds and distributional information.

\subsubsection*{Latent variable representation}\label{sec:latent-variable}

The package also enables direct access to the latent variable representation underlying interval-valued observations. The function \texttt{get\_latent\_var()} retrieves the latent microdata associated with each interval, computing realizations of the latent variables \(U_j\) from observed microdata and their associated interval-valued aggregation. This transformation normalizes the microdata to the interval \([-1,1]\). In order to prevent an accumulation of values at the boundaries, values exactly equal to \(-1\) or \(1\) are set to \texttt{NA}. The function returns a matrix of the normalized microdata, with the same size as the original microdata.

Building on this representation, the function \texttt{get\_latent\_param()} provides a flexible framework for specifying and estimating the distributional characteristics of the latent variables. Depending on the selected modeling scenario, the function returns quantities such as means, second-order moments, or more general dependence measures. These quantities are computed either analytically, based on the specified latent distributions and their parameters, or empirically, using available latent microdata. When the option \texttt{estimate.DistParam} is enabled, the triangular and Beta distribution parameters are estimated from the latent observations. For the triangular distribution, the mode is estimated as three times the sample mean of the latent observations, whereas for the Beta distribution, the shape parameters are estimated by maximum likelihood, using the \texttt{MASS} package \citep{MASS}. Alternatively, when the \texttt{LatentDist} is set to \texttt{"KDE"}, KDE provides a nonparametric estimate of the distribution. The function then computes the required moments and dependence measures according to the assumed distributions, including pairwise measures defined through the integral of the products of the corresponding quantile functions. Currently, the computation of pairwise dependence measures is implemented for combinations of latent variables following triangular or Beta distributions, or characterized using KDE.

\subsection{Implemented methods and functions}\label{sec:implemented-methods}

The \texttt{AIDA} package implements the main methods presented in the previous sections through a collection of functions organized according to their methodological purpose. These functions cover descriptive analysis, distance-based methods, robust estimation and outlier detection, and Shapley-based measures of variable importance and interaction.

\subsubsection*{Descriptive statistics}\label{descriptive-statistics}

Basic descriptive methods for interval-valued data are implemented through the functions \texttt{int\_mean()}, \texttt{int\_cov()}, \texttt{int\_cor()}, \texttt{int\_cov\_z()}, and \texttt{int\_mean\_z()}. The function \texttt{int\_mean()} computes the sample barycenter, while the function \texttt{int\_cov()}/\texttt{int\_cor()} computes the covariance/correlation matrix for interval-valued variables (see equation \eqref{eq:covmatrix}) according to the adopted representation and the corresponding symbolic data model based on the \texttt{intData} object or the sample covariance matrices of the centers, ranges, and cross centers-ranges. Alternatively, the functions \texttt{int\_cov\_z()} and \texttt{int\_mean\_z()} compute, respectively, the sample covariance matrix and mean vector based on an indicator vector \texttt{z}, where a value of one (zero) indicates the corresponding observation is (not) taken into account for the computation. These quantities provide the basis for several of the multivariate methods implemented in the package.

\subsubsection*{\texorpdfstring{Distances for \texttt{intData}}{Distances for intData}}\label{distances-for-intdata}

Distance-based analysis is supported by \texttt{Mallows\_dist()}, which computes the squared Mallows distance (see equation \eqref{eq:mallows}) between interval-valued observations and a location measure, and by the functions \texttt{IMah\_dist()} and \texttt{IMah\_dist\_pairs()}. The function \texttt{IMah\_dist()} computes the squared Interval-Mahalanobis distance of observations from a reference location (see equation \eqref{eq:d2mah}), whereas \texttt{IMah\_dist\_pairs()} computes the corresponding pairwise squared distances. Since the Mallows distance is a particular case of the Interval-Mahalanobis distance, the former can be computed by calling the latter with the identity matrix as the covariance matrix.

\subsubsection*{Robust estimation and outlier detection}\label{robust-estimation-and-outlier-detection}

Robust multivariate analysis and outlier detection are implemented through \texttt{IMCD()}, \texttt{farness()}, and \texttt{int\_outliers()}. The function \texttt{IMCD()} implements the IMCD estimator through an extension of the FastMCD algorithm \citep{fastMCD}, providing robust estimates of location and covariance, and the corresponding robust squared Interval-Mahalanobis distance. The subset size (\texttt{m}) for the IMCD estimator is set as \(75\%\) of the dataset size, as default. This function includes an optional one-step reweighting of the estimates based on a cutoff applied to the computed distances. This cutoff can be chosen from a number of options, with \texttt{farness()} providing the needed transformations for the default case \texttt{"farness"} (adapted from \texttt{cellWise} \citep{cellWise}). Other options include the \(97.5\%\) Chi-squared quantile, adjusted boxplots \citep{adjbox} (using \texttt{robustbase} \citep{robustbase}), and scaled F distribution quantiles \citep{Cerioli} (adapted from \texttt{CerioliOutlierDetection} \citep{CerioliOutlierDetection}). Based on the robust distances, \texttt{int\_outliers()} provides a rule for identifying outlying observations with customizable cutoff criteria (where \texttt{farness()} is called again; see \hyperref[sec:outlier-detection]{Section 2.2}).

\subsubsection*{Shapley value}\label{shapley-value}

The package also implements the explainable outlier detection methods introduced in \hyperref[sec:explainability]{Section 2.3} based on Shapley values. The function \texttt{int\_Shapley()} computes Shapley values for interval-valued data, decomposing the squared Interval-Mahalanobis distance into additive outlyingness contributions of the variables. The function \texttt{int\_Shapley\_decomp()} provides a decomposition of these contributions into centers, ranges, and the interaction between centers and ranges, allowing their different components to be examined separately. Finally, \texttt{int\_Shapley\_interaction()} computes the Shapley interaction index, extending the analysis from individual variable contributions to interactions among pairs of variables.

\subsection{Visualization tools for interval-valued data}\label{sec:visualization-tools}

In addition to methods for constructing and analyzing interval-valued data, \texttt{AIDA} provides a set of visualization tools designed to support exploratory analysis and the interpretation of the package's statistical methods. These functions include graphical representations of interval-valued observations, distance-based diagnostic plots, and visualizations for the interpretation of Shapley values and interaction indices.

\subsubsection*{\texorpdfstring{Exploratory visualization of \texttt{intData}}{Exploratory visualization of intData}}\label{exploratory-visualization-of-intdata}

The functions \texttt{plot\_scatter\_int()} and \texttt{plot\_pairs\_int()} provide graphical tools for exploring the relationships between interval-valued variables. The function \texttt{plot\_scatter\_int()} produces scatter plots designed to represent interval-valued observations, visualizing the intervals as rectangles or crosses. The function \texttt{plot\_pairs\_int()}, in turn, extends the traditional pairs plot to interval-valued data, showing a scatter plot of matrices, with bivariate interval scatter plots below the diagonal, variables' names on the diagonal, and all the symbolic correlations above the diagonal. These plots facilitate the exploration of pairwise relationships while retaining the intervals' information.

\subsubsection*{Distance and outlier diagnostics}\label{distance-and-outlier-diagnostics}

Distance-based methods implemented in the package are complemented by the functions \texttt{plot\_interval\_dist()} and \texttt{plot\_dist\_dist()}. The former plots the Interval-Mahalanobis distances from the observations to the barycenter, providing a graphical tool for assessing their relative departure from the central structure of the data. The latter produces a Distance--Distance plot comparing distances from observations to the barycenter computed using different specifications or estimators. Additionally, \texttt{plot\_dist\_dist()} can generate an interactive \texttt{plotly} \citep{plotly} object, enabling further exploration. Together, these functions support the graphical identification and investigation of potentially atypical observations.

\subsubsection*{Shapley value visualizations}\label{shapley-value-visualizations}

Several graphical representations are available for interpreting the Shapley values computed by \texttt{int\_Shapley()}. The functions \texttt{plot\_bar\_int\_Shapley()}, \texttt{plot\_beeswarm\_int\_Shapley()}, \texttt{plot\_radar\_int\_Shapley()}, and \texttt{plot\_tile\_int\_Shapley()} display Shapley values using, respectively, bar, beeswarm, radar, and tile plots. The availability of different graphical representations allows the contribution of the variables to be examined from multifaceted perspectives. In particular, \texttt{plot\_beeswarm\_int\_Shapley()} optionally creates an interactive \texttt{plotly} object for further exploration. Moreover, since these functions take as argument a matrix containing the Shapley values, their possible use is amplified to any matrix. For example, for the conventional case, one can obtain the Shapley values using the \texttt{ShapleyOutlier} \citep{ShapleyOutlier} package and then take advantage of these functions.

The function \texttt{plot\_bar\_int\_Shapley\_decomp()} further supports the interpretation of the Shapley value decomposition by displaying the contributions associated with centers, ranges, and their cross component. This visualization provides a direct graphical representation of the different sources contributing to the overall importance assigned to each variable. Finally, \texttt{plot\_int\_Shapley\_inter()} provides a graphical representation of the Shapley interaction indices. This function complements the analysis of individual variable contributions by facilitating the interpretation of the interactions between variables.

\subsection{Interval-valued datasets}\label{sec:datasets}

The \texttt{AIDA} package includes four interval-valued datasets from different application domains. Depending on the dataset, both the underlying microdata and different interval representations are available, allowing users to explore the aggregation of microdata into interval-valued observations and the construction of \texttt{intData} objects. These datasets can be used to explore the \texttt{intData} class and its associated methods, as well as to reproduce and adapt the examples provided in the package documentation.

\subsubsection*{\texorpdfstring{\texttt{creditcard}}{creditcard}}\label{creditcard}

The \texttt{creditcard} dataset \citep{billard_diday_2006} contains credit card expenditure data for five features: \emph{Food}, \emph{Social}, \emph{Travel}, \emph{Gas}, and \emph{Clothes}. The data were aggregated by person-month, resulting in \(36\) interval-valued observations. In addition to the underlying microdata, the dataset provides minimum--maximum and center--range representations, together with an \texttt{intData} object. The latter was constructed under the assumption that the underlying microdata follow symmetric triangular distributions (see details about this choice on \citet{oliveira2025}). This dataset provides a compact example illustrating the different representations of interval-valued data supported by \texttt{AIDA}, as illustrated in \hyperref[sec:creditcard]{Section 4.1} and package vignette \href{https://catarinaploureiro.github.io/AIDA/articles/intData_examples.html}{\emph{Class \texttt{intData} examples}}.

\subsubsection*{\texorpdfstring{\texttt{entrecampos\_air\_quality}}{entrecampos\_air\_quality}}\label{entrecampos_air_quality}

The \texttt{entrecampos\_air\_quality} dataset \citep{loureiro2025} contains air pollutant concentration measurements collected at a monitoring station in Entrecampos, Lisbon, during the years 2019--2021. The original data comprise measurements for nine pollutants, with the data aggregated by day to construct interval-valued observations. The dataset includes the raw and logarithmically transformed microdata, daily minimum and maximum values of each pollutant, and a corresponding \texttt{intData} object. Due to the high number of missing values, \emph{benzene} was excluded from the transformed and interval-valued representations, while the remaining missing values were handled through interpolation. The \texttt{intData} object was constructed by estimating the parameters of the latent distributions using KDE. \hyperref[sec:entrecampos]{Section 4.2} exemplifies the methods implemented in the package using this dataset.

\subsubsection*{\texorpdfstring{\texttt{intCars}}{intCars}}\label{intcars}

The \texttt{intCars} dataset (retrieved from the \texttt{MAINT.Data} package) contains interval-valued information on car specifications, aggregated by car model. The original representation provides lower and upper bounds for variables related to \emph{Engine Capacity}, \emph{Top Speed}, \emph{Acceleration}, and \emph{Price}. The dataset also includes an \texttt{intData} object with 27 interval-valued observations and four variables, with price represented on a logarithmic scale. Since the underlying microdata are not available, the corresponding latent distributions were specified using the default parameters under a uniform distribution assumption. Thus, \texttt{intCars} also illustrates the use of the \texttt{intData} class when only interval bounds, rather than the original microdata, are available. This dataset is used in the package's vignettes: \href{https://catarinaploureiro.github.io/AIDA/articles/IMCD_examples.html}{\emph{IMCD estimator examples}} and \href{https://catarinaploureiro.github.io/AIDA/articles/Shapley_examples.html}{\emph{Shapley value examples}}.

\subsubsection*{\texorpdfstring{\texttt{spotify\_tracks}}{spotify\_tracks}}\label{spotify_tracks}

The \texttt{spotify\_tracks} dataset \citep{loureiro2026, loureiro2026b} contains audio features for Spotify tracks, aggregated by track genre. It includes 11 variables describing characteristics: \emph{duration}, \emph{danceability}, \emph{energy}, \emph{loudness}, \emph{speechiness}, \emph{acousticness}, \emph{instrumentalness}, \emph{liveness}, \emph{valence}, \emph{tempo}, and \emph{popularity}. The dataset provides the original and transformed microdata (scaling and logarithmic transformation of some variables), as well as two \texttt{intData} objects containing \(111\) interval-valued observations. These objects were constructed from the transformed microdata using two different aggregation procedures: minimum--maximum aggregation and trimmed aggregation using the \(1\%\) and \(99\%\) quantiles. This dataset is used in the package's vignettes: \href{https://catarinaploureiro.github.io/AIDA/articles/IMCD_examples.html}{\emph{IMCD estimator examples}} and \href{https://catarinaploureiro.github.io/AIDA/articles/Shapley_examples.html}{\emph{Shapley value examples}}.

\section{Examples}\label{sec:examples}

\subsection{Credit card data analysis}\label{sec:creditcard}

To highlight the role of microdata structure in interval-valued data, we consider a simple example adapted from the \texttt{AIDA} package vignette \href{https://catarinaploureiro.github.io/AIDA/articles/intData_examples.html}{\emph{Class \texttt{intData} examples}}. The Credit Card dataset contains information on credit card expenses of three persons over one year \citep{billard_diday_2006}, comprising five variables: \emph{Food}, \emph{Social}, \emph{Travel}, \emph{Gas}, and \emph{Clothes}. The aggregation of the microdata was done by person-month, resulting in \(n=36=3\times 12\) observations. The \texttt{creditcard} dataset is included in \texttt{AIDA} with the following components: \texttt{microdata}, a data frame with \(1000\) rows and seven columns (five variables plus person and month) containing the microdata; \texttt{min\_max}, a data frame with \(36\) rows and \(10\) columns, where each row corresponds to a different interval observation, and columns give the minimum and maximum values for each variable; \texttt{centers\_ranges}, a data frame with \(36\) rows and \(10\) columns, where columns correspond to the centers and ranges of the interval data; \texttt{intData}, an \texttt{intData} object with \(36\) interval-valued observations and five variables, constructed assuming the microdata follow symmetric triangular distributions.

We can exemplify the different possible ways to create an \texttt{intData} object using this dataset, depending on the assumptions about the latent distribution of the microdata. For instance, an \texttt{intData} object can be created using only the \texttt{min\_max} component of the dataset, without information from the microdata (which sometimes is not available). In this case, a continuous uniform distribution is assumed for the latent variables' distribution, which corresponds to the symmetric and identically distributed (i.d.) case (\texttt{LatentCase="U\_id\_symmetric"}), with the \texttt{LatentParam} set as \(\delta = 1/12 \simeq 0.8333\). This is the default setting, given by the following function call.

\begin{Shaded}
\begin{Highlighting}[]
\NormalTok{creditcard\_int\_unif }\OtherTok{\textless{}{-}} \FunctionTok{intData}\NormalTok{(creditcard}\SpecialCharTok{$}\NormalTok{min\_max,}
  \AttributeTok{Seq =} \StringTok{"LbUb\_VarbyVar"}\NormalTok{, }\AttributeTok{VarNames =} \FunctionTok{colnames}\NormalTok{(creditcard}\SpecialCharTok{$}\NormalTok{microdata)[}\DecValTok{3}\SpecialCharTok{:}\DecValTok{7}\NormalTok{]}
\NormalTok{)}
\NormalTok{creditcard\_int\_unif}\SpecialCharTok{@}\NormalTok{LatentParam}
\end{Highlighting}
\end{Shaded}

\begin{verbatim}
## [[1]]
## [1] 0.08333333
\end{verbatim}

Since the microdata are indeed available, we can take a closer look at the distribution of the latent variables. The \texttt{get\_latent\_var()} function can be used to obtain the latent variables observed values, by normalizing the microdata into the \([-1,1]\) interval. In this example, we use the \texttt{min\_max} component of the dataset to normalize the microdata, but we could also use the \texttt{centers\_ranges} component. The aggregation criterion is by month and name, so we create a new variable that combines the name and month to use as the grouping variable for the normalization process, as shown by the following code snippet.

\begin{Shaded}
\begin{Highlighting}[]
\NormalTok{credit\_agrby }\OtherTok{\textless{}{-}} \FunctionTok{factor}\NormalTok{(}\FunctionTok{paste}\NormalTok{(creditcard}\SpecialCharTok{$}\NormalTok{microdata}\SpecialCharTok{$}\NormalTok{Name,}
\NormalTok{  creditcard}\SpecialCharTok{$}\NormalTok{microdata}\SpecialCharTok{$}\NormalTok{Month,}
  \AttributeTok{sep =} \StringTok{"\_"}
\NormalTok{))}
\NormalTok{creditcard\_U }\OtherTok{\textless{}{-}} \FunctionTok{get\_latent\_var}\NormalTok{(}
  \AttributeTok{microdata =}\NormalTok{ creditcard}\SpecialCharTok{$}\NormalTok{microdata[, }\DecValTok{3}\SpecialCharTok{:}\DecValTok{7}\NormalTok{],}
  \AttributeTok{macrodata =}\NormalTok{ creditcard}\SpecialCharTok{$}\NormalTok{min\_max,}
  \AttributeTok{agrlevels =} \FunctionTok{rownames}\NormalTok{(creditcard}\SpecialCharTok{$}\NormalTok{min\_max),}
  \AttributeTok{agrby =}\NormalTok{ credit\_agrby, }\AttributeTok{Seq =} \StringTok{"LbUb\_VarbyVar"}
\NormalTok{)}
\end{Highlighting}
\end{Shaded}

\begin{figure}[ht]
\centering 
\includegraphics[width=1\linewidth]{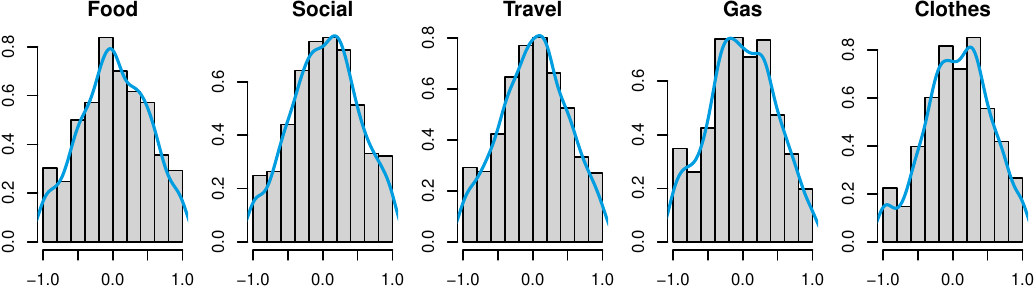}
\caption{Histograms and density plots of the latent variables for the Credit Card dataset, obtained by normalizing the microdata using the min-max values. The distributions appear to be approximately triangular and symmetric.}
\label{fig:creditcard-density}
\end{figure}

Figure \ref{fig:creditcard-density} shows the distribution of the latent variables using histograms and density plots, for the five variables under study. After examining the distribution of the latent variables, we can assume that their distributions are approximately triangular and symmetric. Then, we can create an \texttt{intData} object using the \texttt{min\_max} component of the dataset, specifying the latent distribution as \texttt{"Triang"}, which corresponds to \texttt{LatentCase="U\_id\_symmetric"}, with the \texttt{LatentParam} set as \(\delta = 1/24 \simeq 0.0417\). This is done using the following function call.

\begin{Shaded}
\begin{Highlighting}[]
\NormalTok{creditcard\_int\_triang }\OtherTok{\textless{}{-}} \FunctionTok{intData}\NormalTok{(creditcard}\SpecialCharTok{$}\NormalTok{min\_max,}
  \AttributeTok{Seq =} \StringTok{"LbUb\_VarbyVar"}\NormalTok{, }\AttributeTok{LatentDist =} \StringTok{"Triang"}\NormalTok{,}
  \AttributeTok{VarNames =} \FunctionTok{colnames}\NormalTok{(creditcard}\SpecialCharTok{$}\NormalTok{microdata)[}\DecValTok{3}\SpecialCharTok{:}\DecValTok{7}\NormalTok{]}
\NormalTok{)}
\end{Highlighting}
\end{Shaded}

Alternatively, we can create an \texttt{intData} object using the \texttt{centers\_ranges} component of the dataset. As for the latent variables' distribution parameters, since we have the microdata available, another option is to estimate the parameters directly based on the microdata by setting \texttt{LatentCase\ =\ "General"}, which automatically defaults \texttt{LatentDist} to \texttt{"KDE"}. As the following command illustrates, it is necessary to specify the \texttt{Umicro} argument, which contains the normalized microdata values.

\begin{Shaded}
\begin{Highlighting}[]
\NormalTok{creditcard\_int\_KDE }\OtherTok{\textless{}{-}} \FunctionTok{intData}\NormalTok{(creditcard}\SpecialCharTok{$}\NormalTok{centers\_ranges,}
  \AttributeTok{Seq =} \StringTok{"AllCen\_AllRng"}\NormalTok{, }\AttributeTok{VarNames =} \FunctionTok{colnames}\NormalTok{(creditcard}\SpecialCharTok{$}\NormalTok{microdata)[}\DecValTok{3}\SpecialCharTok{:}\DecValTok{7}\NormalTok{],}
  \AttributeTok{LatentCase =} \StringTok{"General"}\NormalTok{, }\AttributeTok{Umicro =}\NormalTok{ creditcard\_U}
\NormalTok{)}
\end{Highlighting}
\end{Shaded}

In the majority of the cases, however, the macrodata has to be constructed from the microdata. The \texttt{micro2intData()} function can be used to create an \texttt{intData} object from the microdata, by aggregating the microdata according to a specified criterion. Here, as mentioned before, we aggregate the microdata by month and name. We also specify the latent case as \texttt{"General"}, which automatically defaults the latent distribution to \texttt{"KDE"}, in order to estimate the parameters based on the microdata. If the \texttt{LatentCase} argument was omitted, it would assume the latent distribution is uniform (symmetric and i.d.). The \texttt{micro2intData()} can be called as follows, where the \texttt{agrby} argument specifies the grouping variable for aggregation.

\begin{Shaded}
\begin{Highlighting}[]
\NormalTok{creditcard\_int }\OtherTok{\textless{}{-}} \FunctionTok{micro2intData}\NormalTok{(creditcard}\SpecialCharTok{$}\NormalTok{microdata[, }\DecValTok{3}\SpecialCharTok{:}\DecValTok{7}\NormalTok{],}
  \AttributeTok{agrby =}\NormalTok{ credit\_agrby, }\AttributeTok{LatentCase =} \StringTok{"General"}
\NormalTok{)}
\end{Highlighting}
\end{Shaded}

Now, having created the \texttt{intData} object, we can, for example, compute the sample barycenter and the classical symbolic covariance and correlation of the interval data using the following functions, respectively.

\begin{Shaded}
\begin{Highlighting}[]
\NormalTok{creditcard\_mean }\OtherTok{\textless{}{-}} \FunctionTok{int\_mean}\NormalTok{(creditcard\_int)}
\NormalTok{creditcard\_cov }\OtherTok{\textless{}{-}} \FunctionTok{int\_cov}\NormalTok{(creditcard\_int)}
\NormalTok{creditcard\_cor }\OtherTok{\textless{}{-}} \FunctionTok{int\_cor}\NormalTok{(creditcard\_int)}
\end{Highlighting}
\end{Shaded}

\begin{figure}[ht]
\centering 
\includegraphics[width=0.5\linewidth]{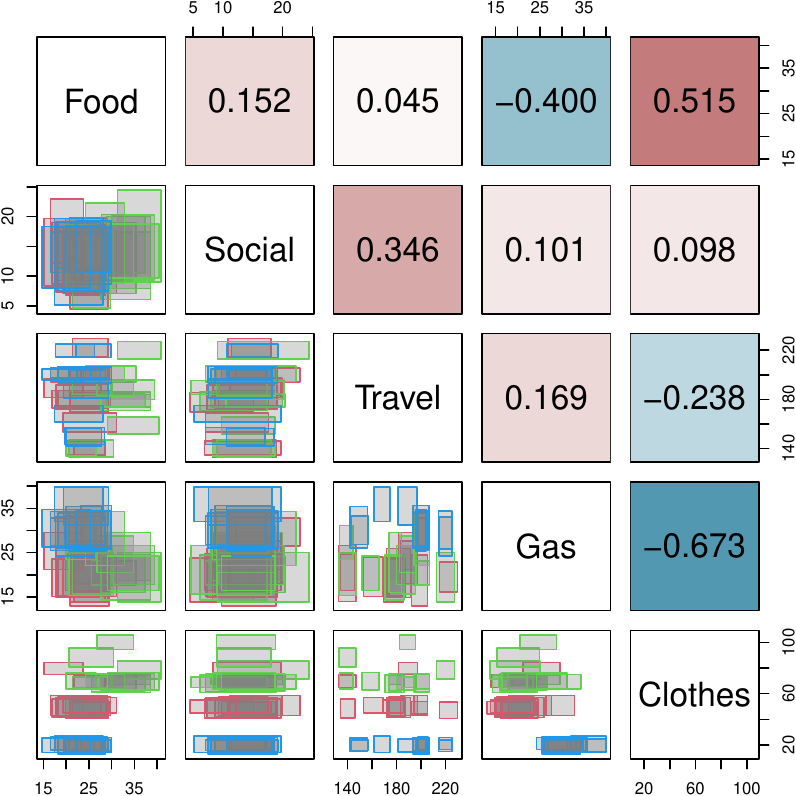} 
\caption{Pairs plot showing scatter plots of the interval-valued data in the lower triangular and the classical interval correlation matrix in the upper triangular. The different people are represented by the colors of the rectangles.}\label{fig:creditcard-pairsplot}
\end{figure}

Additionally, we can visualize the interval data using the \texttt{plot\_pairs\_int()} function (using the command bellow), which creates a pairs plot for interval-valued data, as shown in Figure \ref{fig:creditcard-pairsplot}. The lower triangular shows scatter plots of the variables, while the upper triangular shows the interval correlation matrix. The colors of the scatter plots in the lower triangular correspond to the different people in the dataset, while the colors of the interval correlation matrix in the upper triangular correspond to the correlation values, ranging from blue (negative correlation) to red (positive correlation). We can see that the variables \emph{Food} and \emph{Clothes} are slightly positively correlated, while \emph{Gas} and \emph{Clothes} are slightly negatively correlated, but in general, the variables are not strongly correlated. Furthermore, the scatter plots in the lower triangular show that \emph{Clothes} seems to separate the three people.

\begin{Shaded}
\begin{Highlighting}[]
\NormalTok{colors }\OtherTok{\textless{}{-}} \FunctionTok{c}\NormalTok{(}\FunctionTok{rep}\NormalTok{(}\DecValTok{2}\NormalTok{, }\DecValTok{12}\NormalTok{), }\FunctionTok{rep}\NormalTok{(}\DecValTok{3}\NormalTok{, }\DecValTok{12}\NormalTok{), }\FunctionTok{rep}\NormalTok{(}\DecValTok{4}\NormalTok{, }\DecValTok{12}\NormalTok{))}
\FunctionTok{plot\_pairs\_int}\NormalTok{(creditcard\_int, }\AttributeTok{corr =}\NormalTok{ creditcard\_cor, }\AttributeTok{palette =}\NormalTok{ colors)}
\end{Highlighting}
\end{Shaded}

We can also compute the Mallows distance of each observation to the barycenter using the \texttt{Mallows\_dist()} function directly or the \texttt{IMah\_dist()} function with the identity matrix as the covariance matrix. The function \texttt{IMah\_dist\_pairs()} can also be used to compute the pairwise Mallows distances between all observations using the same logic. The following code snippet illustrates these computations.

\begin{Shaded}
\begin{Highlighting}[]
\NormalTok{I\_p }\OtherTok{\textless{}{-}} \FunctionTok{diag}\NormalTok{(creditcard\_int}\SpecialCharTok{@}\NormalTok{NIVar)}
\NormalTok{creditcard\_dM }\OtherTok{\textless{}{-}} \FunctionTok{Mallows\_dist}\NormalTok{(creditcard\_int)}
\NormalTok{creditcard\_dM2 }\OtherTok{\textless{}{-}} \FunctionTok{IMah\_dist}\NormalTok{(creditcard\_int, }\AttributeTok{cov =}\NormalTok{ I\_p, }\AttributeTok{mean =}\NormalTok{ creditcard\_mean)}
\NormalTok{creditcard\_pairs }\OtherTok{\textless{}{-}} \FunctionTok{IMah\_dist\_pairs}\NormalTok{(creditcard\_int, }\AttributeTok{cov =}\NormalTok{ I\_p)}

\FunctionTok{all.equal}\NormalTok{(creditcard\_dM, creditcard\_dM2)}
\end{Highlighting}
\end{Shaded}

\begin{verbatim}
## [1] TRUE
\end{verbatim}

\subsection{Air quality data analysis}\label{sec:entrecampos}

The Entrecampos Air Quality dataset \citep{qualar} was collected from an air quality monitoring station in Entrecampos, Lisbon, Portugal. This monitoring station registers the hourly concentration of nine pollutants (\(\mu g/m^3\)): sulfur dioxide (SO\(_2\)), particles \(<10 \mu m\), ozone (O\(_3\)), nitrogen dioxide (NO\(_2\)), carbon monoxide (CO), benzene (C\(_6\)H\(_6\)), particles \(<2.5 \mu m\), nitrogen oxides (NO\(_x\)), and nitrogen monoxide (NO). For more details on this dataset, see \citet{loureiro2025}.

The dataset \texttt{entrecampos\_air\_quality} included in \texttt{AIDA} covers the period from January 1, 2019, to December 31, 2021, with the aggregation of the microdata having been done by day. It is comprised by the components: \texttt{microdata\_raw}, a data frame with \(26304\) rows and \(11\) columns, containing the raw microdata; \texttt{microdata\_transformed}, a data frame with \(26304\) rows and \(10\) columns, containing the microdata, where logarithmic transformations were applied to all variables; \texttt{min\_max}, a data frame with \(1096\) rows and \(17\) columns, where each row corresponds to a different day, and columns give the minimum and maximum values for each variable, with interpolation having been applied to deal with missing values; \texttt{intData}, an \texttt{intData} object with \(1096\) observations (days), eight variables, and \(24\) observations per day (\texttt{NbMicroUnits}), constructed using KDE for estimating the parameters of the latent distributions. For the \texttt{microdata\_transformed}, \texttt{min\_max}, and \texttt{intData} components, the pollutant \emph{benzene} was removed due to a high number of missing values.

In this example, we restrict our analysis to the year \(2019\), for brevity and ease of visualization of the plots. Therefore, we create a new \texttt{intData} object based solely on the measurements from \(2019\), estimating the latent distribution parameters based on the microdata through KDE. This new dataset is composed of \(292\) interval-valued observations and eight variables, with \(73\) interval-valued observations having been removed since at least one variable was missing. This \texttt{intData} object can be created as follows.

\begin{Shaded}
\begin{Highlighting}[]
\NormalTok{entrecampos }\OtherTok{\textless{}{-}}\NormalTok{ entrecampos\_air\_quality}\SpecialCharTok{$}\NormalTok{microdata\_transformed}
\NormalTok{entrecampos\_2019 }\OtherTok{\textless{}{-}}\NormalTok{ entrecampos[}\FunctionTok{format}\NormalTok{(entrecampos}\SpecialCharTok{$}\NormalTok{Day, }\StringTok{"\%Y"}\NormalTok{) }\SpecialCharTok{==} \DecValTok{2019}\NormalTok{, ]}

\NormalTok{entrecampos\_int }\OtherTok{\textless{}{-}} \FunctionTok{micro2intData}\NormalTok{(entrecampos\_2019[, }\DecValTok{3}\SpecialCharTok{:}\DecValTok{10}\NormalTok{],}
  \AttributeTok{agrby =} \FunctionTok{factor}\NormalTok{(entrecampos\_2019[, }\DecValTok{2}\NormalTok{]),}
  \AttributeTok{LatentCase =} \StringTok{"General"}
\NormalTok{)}
\end{Highlighting}
\end{Shaded}

Once we have the \texttt{intData} object, we can use the \texttt{IMCD()} function to obtain the IMCD estimates of the barycenter and symbolic covariance matrix, as well as the robust squared Interval-Mahalanobis distances of each observation from said barycenter. In this case, we use the farness cutoff \citep{classmap} with a \(0.975\) threshold (\texttt{cutoff\_lvl}) for the reweighting of the estimates. Next, we can apply the outlier detection rule using the \texttt{int\_outliers()} function, once again with a \(0.975\) farness cutoff (\texttt{cutoff\_lvl}), as follows.

\begin{Shaded}
\begin{Highlighting}[]
\NormalTok{entrecampos\_IMCD }\OtherTok{\textless{}{-}} \FunctionTok{IMCD}\NormalTok{(entrecampos\_int, }\AttributeTok{cutoff\_lvl =} \FloatTok{0.975}\NormalTok{)}
\NormalTok{entrecampos\_out }\OtherTok{\textless{}{-}} \FunctionTok{int\_outliers}\NormalTok{(entrecampos\_IMCD}\SpecialCharTok{$}\NormalTok{robust\_dist, }\AttributeTok{cutoff\_lvl =} \FloatTok{0.975}\NormalTok{)}
\end{Highlighting}
\end{Shaded}

For comparison, we can also compute the classical (non-robust) Interval-Mahalanobis distance using \texttt{IMah\_dist()} and reapply the outlier detection rule with a \(1.5\) adjusted boxplot \citep{adjbox} cutoff (the default \texttt{cutoff\_lvl} for \texttt{cutoff\ =\ "adjbox"}):

\begin{Shaded}
\begin{Highlighting}[]
\NormalTok{entrecampos\_class\_d }\OtherTok{\textless{}{-}} \FunctionTok{IMah\_dist}\NormalTok{(entrecampos\_int, }\AttributeTok{z =} \FunctionTok{rep}\NormalTok{(}\DecValTok{1}\NormalTok{, entrecampos\_int}\SpecialCharTok{@}\NormalTok{NObs))}
\NormalTok{entrecampos\_class\_out }\OtherTok{\textless{}{-}} \FunctionTok{int\_outliers}\NormalTok{(entrecampos\_class\_d, }\AttributeTok{cutoff =} \StringTok{"adjbox"}\NormalTok{)}
\end{Highlighting}
\end{Shaded}

We can now obtain the Shapley value decomposition of the robust squared Interval-Mahalanobis distances obtained above, using the following \texttt{int\_Shapley()} function call. A more detailed decomposition into centers, ranges, and cross centers-ranges, can be computed with the following \texttt{int\_Shapley\_decomp()} function call.

\begin{Shaded}
\begin{Highlighting}[]
\NormalTok{entrecampos\_sh }\OtherTok{\textless{}{-}} \FunctionTok{int\_Shapley}\NormalTok{(entrecampos\_int,}
  \AttributeTok{mean\_c =}\NormalTok{ entrecampos\_IMCD}\SpecialCharTok{$}\NormalTok{mean\_IMCD\_c,}
  \AttributeTok{mean\_r =}\NormalTok{ entrecampos\_IMCD}\SpecialCharTok{$}\NormalTok{mean\_IMCD\_r,}
  \AttributeTok{cov =}\NormalTok{ entrecampos\_IMCD}\SpecialCharTok{$}\NormalTok{cov\_IMCD}
\NormalTok{)}
\NormalTok{entrecampos\_sh\_decomp }\OtherTok{\textless{}{-}} \FunctionTok{int\_Shapley\_decomp}\NormalTok{(entrecampos\_int,}
  \AttributeTok{mean\_c =}\NormalTok{ entrecampos\_IMCD}\SpecialCharTok{$}\NormalTok{mean\_IMCD\_c,}
  \AttributeTok{mean\_r =}\NormalTok{ entrecampos\_IMCD}\SpecialCharTok{$}\NormalTok{mean\_IMCD\_r,}
  \AttributeTok{cov =}\NormalTok{ entrecampos\_IMCD}\SpecialCharTok{$}\NormalTok{cov\_IMCD}
\NormalTok{)}
\end{Highlighting}
\end{Shaded}

In addition, the following \texttt{int\_Shapley\_interaction()} function call provides the Shapley index interaction for each observation.

\begin{Shaded}
\begin{Highlighting}[]
\NormalTok{entrecampos\_sh\_inter }\OtherTok{\textless{}{-}} \FunctionTok{int\_Shapley\_interaction}\NormalTok{(entrecampos\_int,}
  \AttributeTok{mean\_c =}\NormalTok{ entrecampos\_IMCD}\SpecialCharTok{$}\NormalTok{mean\_IMCD\_c,}
  \AttributeTok{mean\_r =}\NormalTok{ entrecampos\_IMCD}\SpecialCharTok{$}\NormalTok{mean\_IMCD\_r,}
  \AttributeTok{cov =}\NormalTok{ entrecampos\_IMCD}\SpecialCharTok{$}\NormalTok{cov\_IMCD}
\NormalTok{)}
\end{Highlighting}
\end{Shaded}

\begin{figure}[ht]
\centering 
\subfloat[\label{fig:entrecampos-plot-1}]{\includegraphics[width=0.49\linewidth]{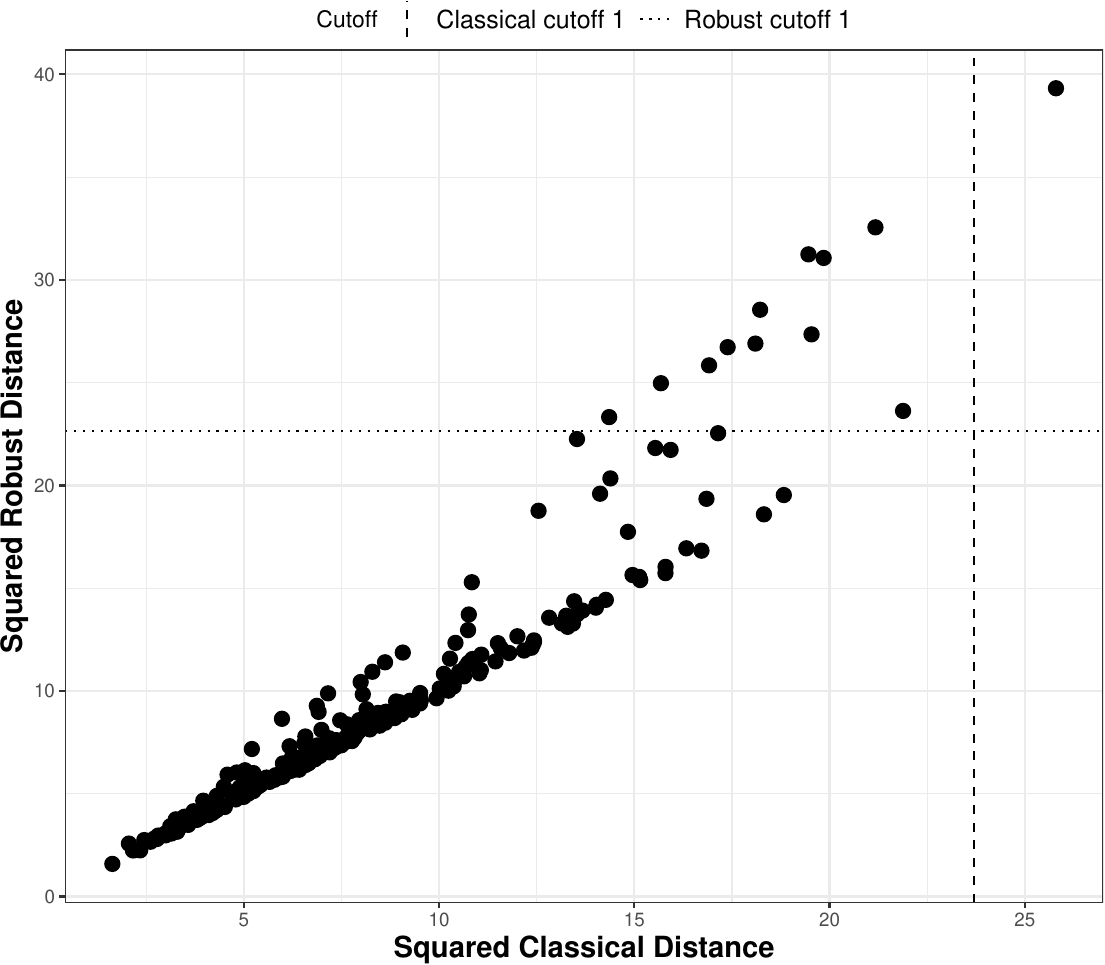}}
\subfloat[\label{fig:entrecampos-plot-2}]{\includegraphics[width=0.49\linewidth]{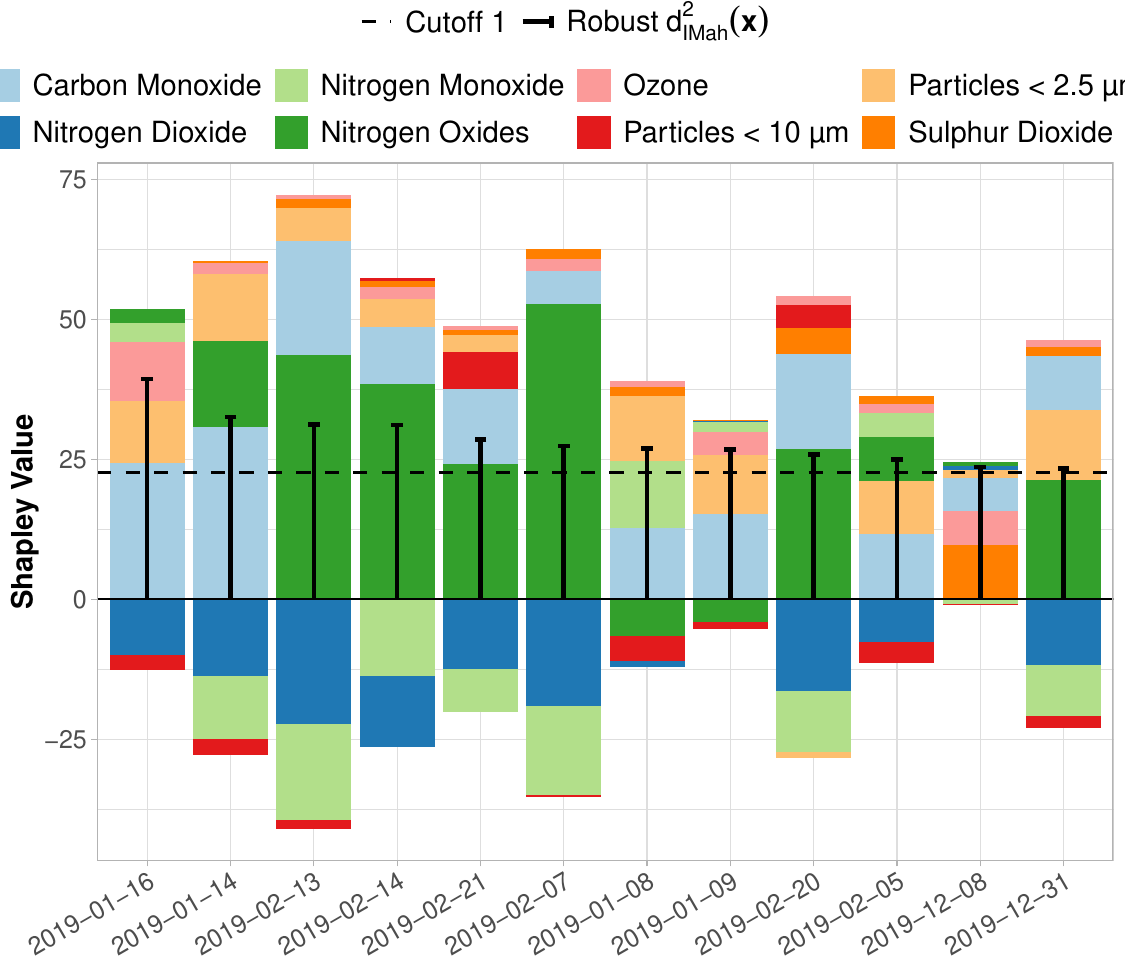} }
\caption{On the left, distance-distance plot (a) of the classical versus robust squared Interval-Mahalanobis distances for the Entrecampos dataset. The horizontal and vertical lines correspond to the 0.975 farness cutoff value and the 1.5 adjusted boxplot cutoff value, respectively. On the right, a barplot (b) of the Shapley value decomposition of the robust squared Interval-Mahalanobis distances. The 12 outliers are shown, ordered from greatest to smallest distance represented by the black vertical lines, with the dashed horizontal line representing the 0.975 farness cutoff.}\label{fig:entrecampos-plot}
\end{figure}

In order to visualize the results obtained, we start by plotting the classical (non-robust) squared Interval-Mahalanobis distances versus their robust counterparts, through the \texttt{plot\_dist\_dist()} function call bellow, which produces Figure \ref{fig:entrecampos-plot}(a). We can see that \(12\) observations lie above the horizontal robust cutoff, corresponding to the detected outliers. Only one observation exceeds the vertical classical cutoff, and it is also identified as an outlier by the robust method. This plot shows the advantage of using the robust distance for outlier detection, since the classical strategy only detects the most extreme outlier.

\begin{Shaded}
\begin{Highlighting}[]
\FunctionTok{plot\_dist\_dist}\NormalTok{(}
  \AttributeTok{class\_dist =}\NormalTok{ entrecampos\_class\_d,}
  \AttributeTok{class\_cutoff =}\NormalTok{ entrecampos\_class\_out}\SpecialCharTok{$}\NormalTok{cutoff\_value[[}\DecValTok{2}\NormalTok{]],}
  \AttributeTok{rob\_dist =}\NormalTok{ entrecampos\_IMCD}\SpecialCharTok{$}\NormalTok{robust\_dist,}
  \AttributeTok{rob\_cutoff =}\NormalTok{ entrecampos\_out}\SpecialCharTok{$}\NormalTok{cutoff\_value}
\NormalTok{)}
\FunctionTok{plot\_bar\_int\_Shapley}\NormalTok{(entrecampos\_sh[entrecampos\_out}\SpecialCharTok{$}\NormalTok{outliers\_names, ],}
  \AttributeTok{cutoff\_value =}\NormalTok{ entrecampos\_out}\SpecialCharTok{$}\NormalTok{cutoff\_value}
\NormalTok{)}
\end{Highlighting}
\end{Shaded}

For the Shapley values, the \texttt{plot\_bar\_int\_Shapley()} call above provides a barplot of the Shapley value decomposition of the robust squared Interval-Mahalanobis distances for the \(12\) outliers detected, as in Figure \ref{fig:entrecampos-plot}(b). We see that the variables that most positively contribute are \emph{Carbon Monoxide} and \emph{Nitrogen Oxides}, while the ones that most negatively contribute are the \emph{Nitrogen Dioxide} and the \emph{Nitrogen Monoxide}. These are pollutants whose major sources are, for instance, energy production and road transport, which aligns with the winter period of the outliers detected for having atypical values.

\begin{figure}[ht]
\centering
\subfloat[\label{fig:entrecampos-shapley-plot-1}]{\includegraphics[width=0.49\linewidth]{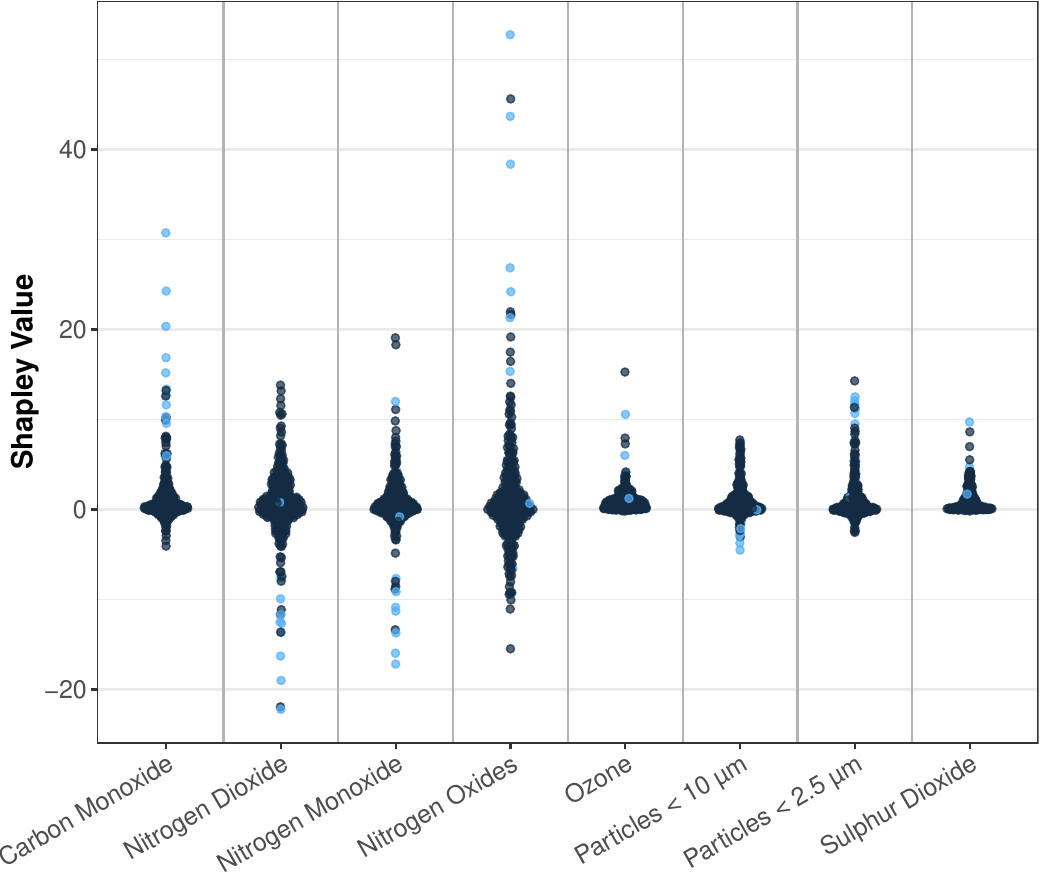}}
\subfloat[\label{fig:entrecampos-shapley-plot-2}]{\includegraphics[width=0.49\linewidth]{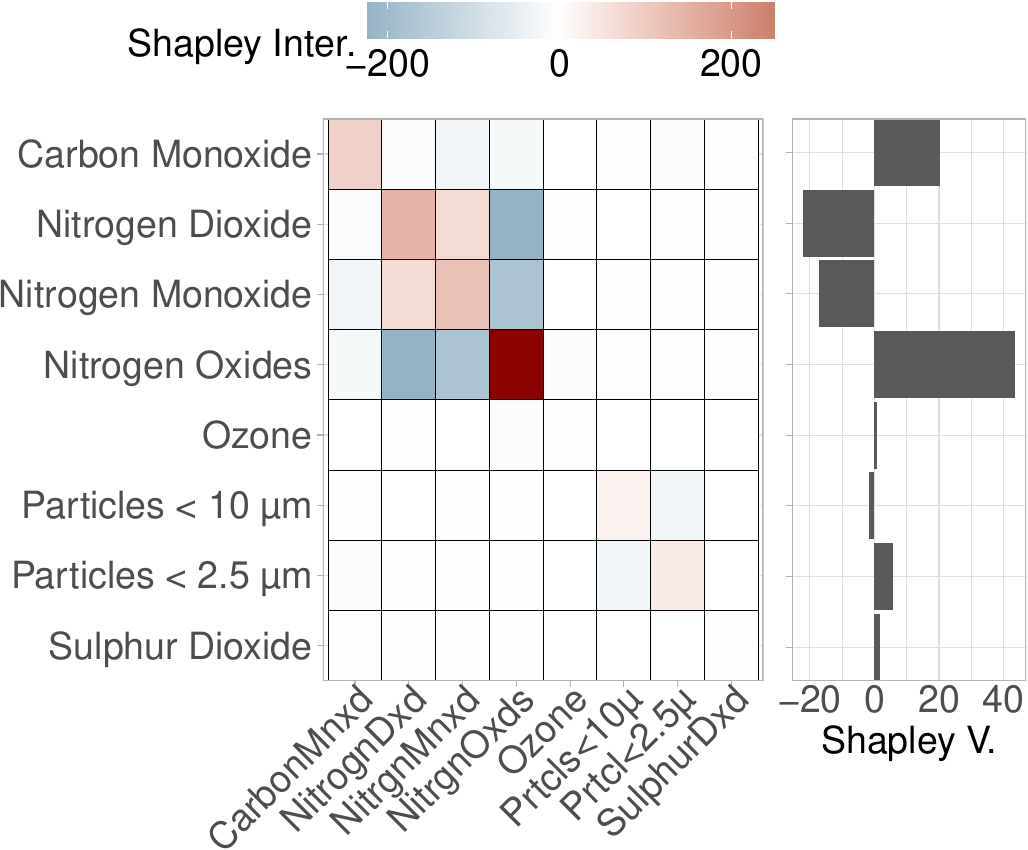} }
\caption{On the left, beeswarm plot (a) of the Shapley value decomposition of the robust squared Interval-Mahalanobis distances for the Entrecampos dataset. The outlying observations are marked in blue. On the right, tile plot (b) of the Shapley interaction index for '2019-02-13', with the corresponding Shapley values on the right.}\label{fig:entrecampos-shapley-plot}
\end{figure}

Figure \ref{fig:entrecampos-shapley-plot}(a) shows a beeswarm plot of the Shapley values obtained from the following \texttt{plot\_beeswarm\_int\_Shapley()} function call, with outliers highlighted in blue. This plot can be used to find potential cellwise outliers. For instance, we see that, for some variables, the observations with the largest positive Shapley values are not outliers in the global sense. After further investigation into these observations, using the \texttt{ggplotly=TRUE} option of the plotting function, the days ``2019-05-12'' and ``2019-08-11'' are potential cellwise outliers for \emph{Nitrogen Monoxide}, as ``2019-11-20'' is for \emph{Ozone}, or ``2019-05-10'' is for \emph{Particles} \(<2.5\mu m\).

\begin{Shaded}
\begin{Highlighting}[]
\FunctionTok{plot\_beeswarm\_int\_Shapley}\NormalTok{(entrecampos\_sh, entrecampos\_out}\SpecialCharTok{$}\NormalTok{is\_outlier }\SpecialCharTok{+} \DecValTok{1}\NormalTok{)}
\FunctionTok{plot\_int\_Shapley\_inter}\NormalTok{(entrecampos\_sh\_inter[[}\StringTok{"2019{-}02{-}13"}\NormalTok{]])}
\end{Highlighting}
\end{Shaded}

The Shapley interaction index can be represented through a tile plot, using the \texttt{plot\_int\_Shapley\_inter()} call above. Figure \ref{fig:entrecampos-shapley-plot}(b) shows the resulting plot for the day ``2019-02-13'', where we see a block of colored tiles for the \emph{Carbon Monoxide} and Nitrogen related variables, and another much lighter one for the particles variables. The rest of the tiles are mostly white. The diagonal values represent the individual contributions of each variable, while the off-diagonal elements represent the joint contributions of pairs of variables. In particular, the individual contribution of \emph{Nitrogen Oxides} is the largest positive contribution, and even though the joint contributions with the other Nitrogen variables are negative, they are not enough to stabilize the outlying effect of \emph{Nitrogen Oxides}. Conversely, \emph{Nitrogen Dioxide} and \emph{Nitrogen Monoxide} have individual and joint positive contributions, but the negative joint contribution with \emph{Nitrogen Oxides} are significant resulting in a stabilizing effect of these two variables to the outlyingness of ``2019-02-13''.

\begin{figure}[ht]
\centering \includegraphics[width=1\linewidth]{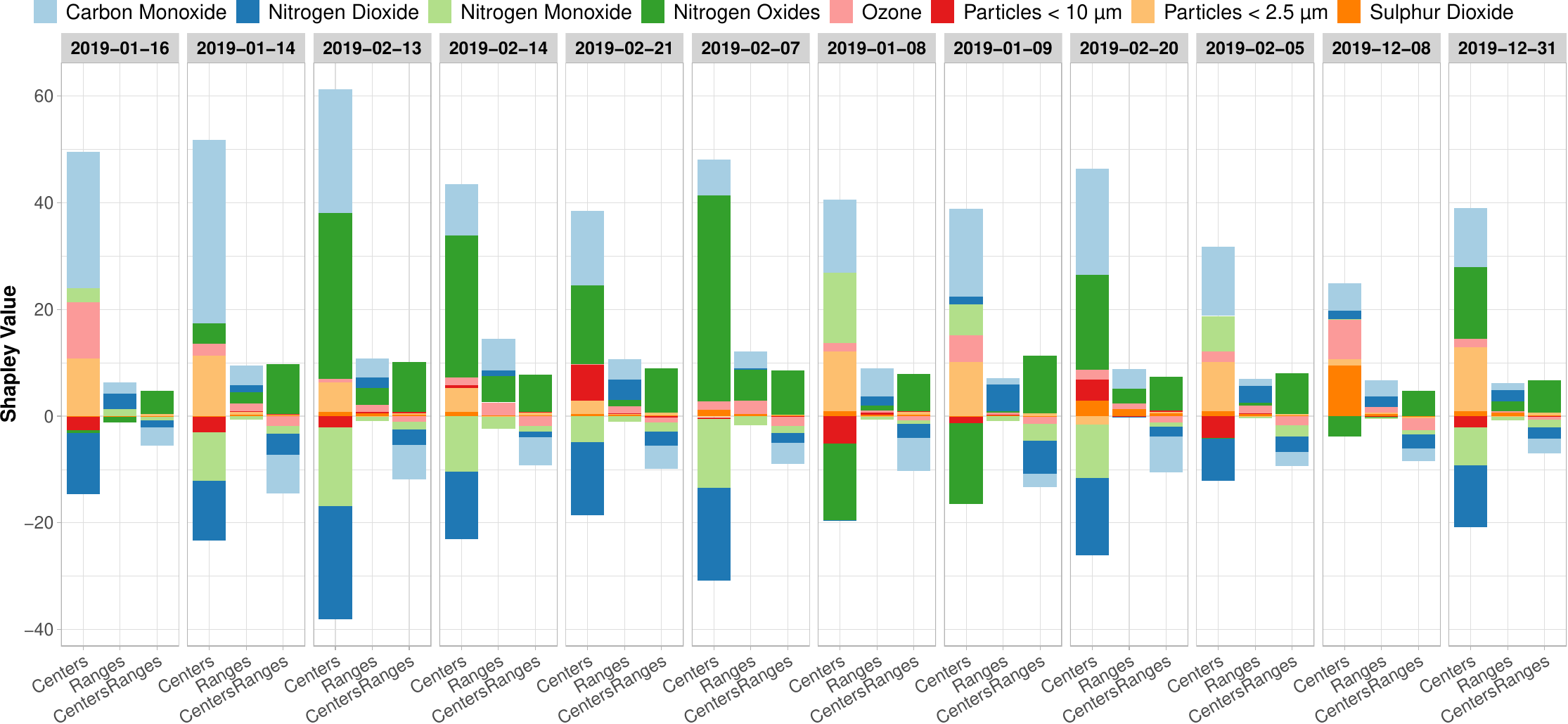} 
\caption{Barplot of the decomposition of the Shapley values by centers, ranges, and cross centers-ranges for the 12 outliers in the Entrecampos dataset.}\label{fig:entrecampos-shapley-decomp-plot}
\end{figure}

Finally, \texttt{plot\_bar\_int\_Shapley\_decomp()} provides a barplot of the decomposition of the Shapley values by centers, ranges, and cross centers-ranges for the \(12\) outliers identified, shown in Figure \ref{fig:entrecampos-shapley-decomp-plot}. We see that the centers are the greatest contributors to the robust squared distance, with the ranges and cross centers-ranges having a smaller part. This is to be expected based on the Shapley value expression. For example, \emph{Carbon Monoxide} and \emph{Nitrogen Dioxide} seem to always have negative values for the cross centers-ranges term. In fact, \emph{Carbon Monoxide} and \emph{Nitrogen Oxides} are the variables that most drive the outlyingness, while \emph{Nitrogen Dioxide} and \emph{Nitrogen Monoxide} are the ones that most have a stabilizing effect.

\begin{Shaded}
\begin{Highlighting}[]
\FunctionTok{plot\_bar\_int\_Shapley\_decomp}\NormalTok{(entrecampos\_sh\_decomp[entrecampos\_out}\SpecialCharTok{$}\NormalTok{outliers\_names])}
\end{Highlighting}
\end{Shaded}

\section{Summary}\label{sec:summary}

\texttt{AIDA} is a package for the analysis of interval-valued data in R, supporting a complete workflow, including data construction, descriptive analysis, distance-based measures, robust estimation, and explainable outlier detection. Built around the \texttt{intData} class, the package represents interval-valued observations through their centers and ranges while retaining information on the underlying latent microdata distribution. A key contribution is the implementation of the IMCD estimator and the corresponding robust Interval-Mahalanobis distance, providing a robust framework for identifying atypical observations. These methods are complemented by Shapley value-based explainability tools, including interaction indices, which decompose the outlyingness measure into variable-specific contributions.

The package also provides visualization tools and example datasets to facilitate exploratory analysis, graphical diagnostics, and interpretation of results. In particular, the plotting functions have a broader scope, allowing the exploration of Shapley values in general, with the beeswarm plot's interactivity enabling a deeper investigation. In addition, we contribute with two new datasets, namely \texttt{entrecampos\_air\_quality} and \texttt{spotify\_tracks}. Together, these features provide a coherent and extensible framework that combines robust methodology with practical tools for the analysis of symbolic interval data.

The package documentation is also available online as a \texttt{pkgdown} website hosted on \href{https://catarinaploureiro.github.io/AIDA/}{GitHub Pages}. The package is currently in version \(0.2.2\), with further methods and features planned to extend its support for interval-valued data analysis.

\section{Acknowledgements}\label{acknowledgements}

Generative AI (GPT-5) was used for language editing and assistance with R coding and troubleshooting, with the authors reviewing and editing the content as needed, and taking full responsibility for the published article.
This work was supported by FCT --- Fundação para a Ciência e Tecnologia, I.P. through grant \href{https://doi.org/10.54499/UI/BD/153720/2021}{UI/\discretionary{}{}{}BD/\discretionary{}{}{}153720/\discretionary{}{}{}2021}, and projects \href{https://doi.org/10.54499/UID/PRR/04621/2025}{UID/\discretionary{}{}{}04621/\discretionary{}{}{}2025} and \href{https://doi.org/10.54499/UID/50014/2025}{UID/\discretionary{}{}{}50014/\discretionary{}{}{}2025}.

  \bibliography{AIDA.bib}

\end{document}